\documentclass[journal,10pt]{IEEEtran}
\usepackage{amsmath, nccmath}
\usepackage{cases} 
\usepackage{amssymb}
\usepackage{cite}
\usepackage{url}
\usepackage{xcolor}
\usepackage{cite,graphicx}
\usepackage{subfigure}
\usepackage{citesort}
\usepackage{fancyhdr}
\usepackage{mdwmath}
\usepackage{mdwtab}
\usepackage{caption}
\usepackage{amsthm}
\usepackage{algorithmic}
\usepackage{algorithm}
\usepackage{threeparttable}
\usepackage{makecell}
\usepackage{booktabs}
\usepackage{pifont}

\newtheorem{theorem}{Theorem}

\newtheorem{lemma}{Lemma}

\newtheorem{corollary}{Corollary}

\makeatletter
\def\ScaleIfNeeded{%
\ifdim\Gin@nat@width>\linewidth \linewidth \else \Gin@nat@width
\fi } \makeatother

\begin{document}

\title{Pinching Antenna Systems (PASS) for Cell-Free Communications}

\author{ 
        {Haochen~Li},~\IEEEmembership{Member,~IEEE,} 
        {Jiayi~Lei},~\IEEEmembership{Member,~IEEE,}
        {Cheng~Zeng},~\IEEEmembership{Member,~IEEE,}  
        {Zhaoming~Hu},~\IEEEmembership{Member,~IEEE,}
        {Chao~Dong},~\IEEEmembership{Senior Member,~IEEE,}
        {Yuanwei~Liu},~\IEEEmembership{Fellow,~IEEE}   
\vspace{-0.85cm}
\thanks{Haochen~Li and Chao~Dong are with the College of Electronic and Information Engineering, Nanjing University of Aeronautics and Astronautics, Nanjing 211106, China (email: haochen.li@nuaa.edu.cn, dch@nuaa.edu.cn).}
\thanks{Jiayi~Lei is with the College of Information Science and Engineering, Hohai University, Nanjing 213200, China (e-mail: jiayi.lei@hhu.edu.cn).}
\thanks{Cheng~Zeng is with the School of Electronic and Optical Engineering, Nanjing University of Science and Technology, Nanjing 210094, China (e-mail: czeng@njust.edu.cn).}
\thanks{Zhaoming~Hu is with School of Computer Science and Technology (School of Big Data), Taiyuan University of Technology, Taiyuan 030002, China (email: huzhaoming@tyut.edu.cn).}
\thanks{Yuanwei Liu is with the Department of Electrical and Electronic Engineering, The University of Hong Kong, Hong Kong. (e-mail: yuanwei@hku.hk).}
}
\maketitle
\begin{abstract}
A pinching antenna system (PASS) assisted cell-free communication system is proposed. A sum rate maximization problem under the BS power budget constraint and the pinching antenna (PA) deployment constraint is formulated. To tackle the proposed non-convex optimization problem, an alternating optimization (AO) algorithm is developed. In particular, the digital beamforming sub-problem is solved using the weighted minimum mean square error (WMMSE) method, whereas the pinching beamforming sub-problem is handled via a penalty based approach combined with element-wise optimization. Simulation results demonstrate that: 1) the PASS assisted cell-free systems achieve superior performance over its multiple-input and multiple-output (MIMO) counterpart; 2) increasing the number of PAs per waveguides can improve the advantage of PASS assisted cell-free systems; and 3) the cell-free architecture mitigates the average user rate degradation as the number of users increases.


\end{abstract}


\vspace{-0.4cm}
\section{Introduction}\label{Introduction}\vspace{-0.1cm}

Recently, the pinching antenna system (PASS) has been proposed as a highly promising antenna architecture, offering a novel system design paradigm~\cite{11222687}. {By enabling the relocation of pinching antennas (PAs) along dielectric waveguides, the PASS offers several key advantages compared with conventional fixed antenna architectures: it enables flexible antenna deployment adaptable to dynamic user distributions, increases beamforming freedom for enhanced interference management, and improves signal strength with decreased signal propagation distance~\cite{11123791}. }

Though the PASS offers the aforementioned advantages for wireless communication systems, how to effectively combine it with existing communication architectures remains an open problem~\cite{10945421}. {In particular, deploying PASS base stations (BSs) within a conventional cellular architecture may exacerbate inter-cell interference. This is because when PAs are positioned closer to the users in their serving cell, they may also become geographically closer to users in neighboring cells. For a similar reason, pilot contamination is also aggravated in PASS assisted cellular communication systems during the channel estimation stage.} These drawbacks lead to a degradation of communication performance and pose new challenges for system design.
This challenging issue can be effectively addressed by adopting the cell-free communication architecture~\cite{ngo2017cell,bjornson2020scalable}. {The cell-free architecture mitigates inter-cell interference through cooperative transmission among BSs. Instead of treating signals from neighboring BSs as interference, multiple cell-free BSs jointly serve users with coordinated beamforming. Consequently, a portion of the inter-cell interference can be transformed into useful cooperative signals. Moreover, coordinated beamforming across PASS BSs enables more effective spatial interference suppression, since both beamforming and antenna positions can be jointly optimized to control the spatial distribution of transmitted energy.}


{Adopting other flexible antenna structures, works~\cite{11018493} and~\cite{10967080} integrate the cell-free architecture into fluid antenna systems and movable antenna systems. These approaches remain susceptible to LoS blockages, since adjusting antennas within a limited wavelength-scale movement region cannot reliably establish unobstructed LoS links. By contrast, PASS enables antenna repositioning over meter-scale waveguides, thereby providing substantially enhanced capability for blockage avoidance and channel reconstruction. Inspired by the concept of distributed antenna systems, work~\cite{11355743} investigates joint transmission strategies between a BS and pinching antennas to improve user performance. Nevertheless, the considered system only focuses on a single-user scenario and does not address the cooperative multi-user transmission and interference management issues faced by cell-free systems.}


Motivated by these observations, {this work introduces a cell-free architecture into PASS assisted communication systems. The main contributions of this work are summarized as follows:}
\begin{itemize}
        \item {A PASS assisted cell-free system is proposed, where the cell-free architecture is leveraged to mitigate the intensifying inter-cell interference that arises from the decreasing distance between antennas and users in PASS assisted communications.} A sum rate maximization problem is formulated under the BS power constraint and PA deployment requirement to jointly optimize the digital beamforming and PA deployment, i.e., the pinching beamforming.
        \item The proposed sum rate maximization problem is converted  into an equivalent problem of minimizing a weighted mean square error using the weight minimum mean square error (WMMSE) method. Then, an alternating optimization (AO) algorithm is proposed to tackle the reformulated problem with penalty method and element-wise optimization.   Semi-closed solutions for beamforming design can be obtained with the proposed algorithm.  
        \item Simulation results verify the superiority of the proposed PASS assisted cell-free system over benchmark schemes, and further reveal the performance benefits brought by increasing the number of PAs per waveguide and by the PASS cell-free architecture in supporting increased user number.
\end{itemize}\vspace{-0.2cm}
\section{System Model}\label{System}
\begin{figure} [htbp]
\centering\vspace{-0.2cm}
\includegraphics[width=0.5\textwidth]{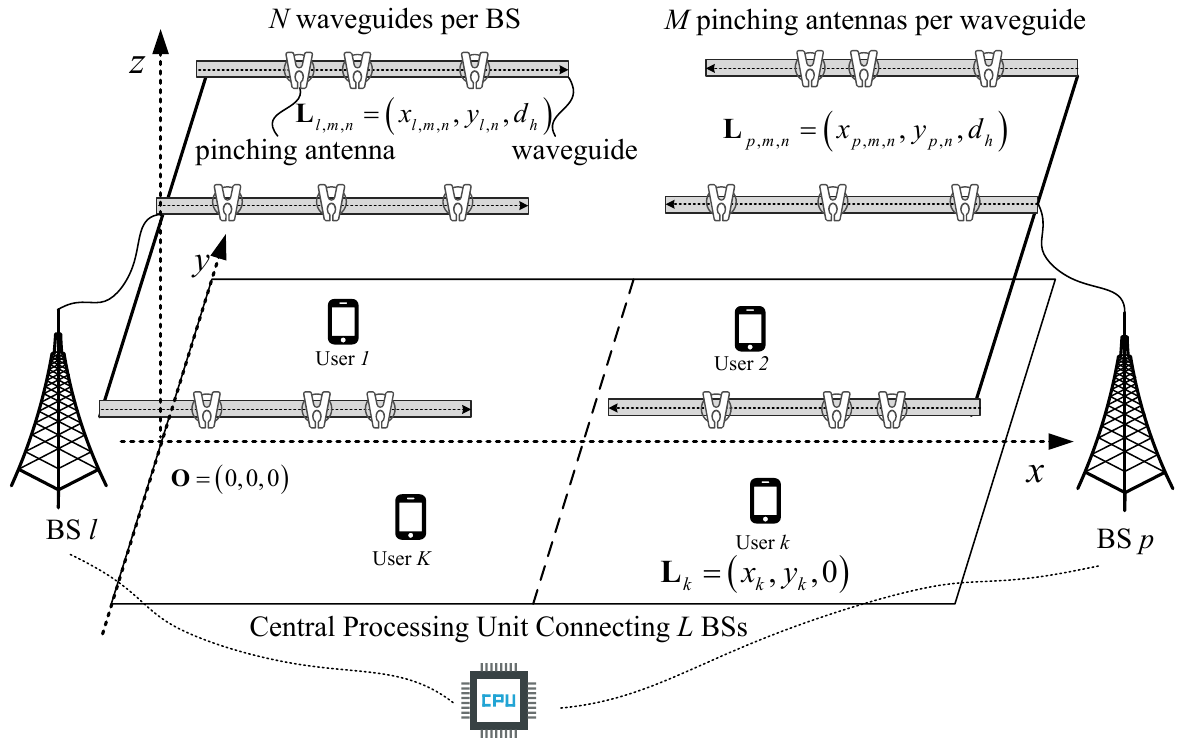}
 \caption{{The proposed PASS assisted cell-free communication system.}\vspace{-0.3cm}
  }
 \label{system_model}
\end{figure}
As depicted in Fig.~\ref{system_model}, a PASS assisted cell-free communication system is investigated in this work, where $K$ communication users are jointly served by $L$ distributed PASS BSs connected to a central processing unit (CPU) for control and planning.  Each PASS BS is equipped with $N$ waveguides on which $M$ PAs can be applied to tunable location. It is assumed that $K\le N$. Let~$\mathcal{L}$,~$\mathcal{N}$, $\mathcal{M}$, and $\mathcal{K}$ denote the index sets of BSs, waveguides in each PASS BS, PAs on each waveguide, and communication users, respectively. All BSs and users are distributed in a service area spanned by $D_x\times D_y$. The location of user $k$ is $\mathbf{L}_{u,k}=\left[x_{u,k},y_{u,k},0\right]^\mathrm{T}, \forall k \in \mathcal{K}$. All waveguides are set parallel to the $x$-axis at the height of $d_h$. The feed point of the $n$-th waveguide of the $l$-th BS is $\mathbf{L}_{l,n}=\left[x_{l,n},y_{l,n},d_h\right]^\mathrm{T}, \forall l \in \mathcal{L}, \forall n \in \mathcal{N}$, while the location of the $m$-th PA on this waveguide is $\mathbf{L}_{l,m,n}=\left[x_{l,m,n},y_l,d_h\right]^\mathrm{T}, \forall l \in \mathcal{L}, \forall m \in \mathcal{M}, \forall n \in \mathcal{N}$. {Through designing the PA deployment variables $x_{l,m,n}$, the PA is moved along the waveguide to jointly customize the communication channels for users. Constrained by the limited length of waveguides, the PA deployment should satisfy\vspace{-0.1cm}
\begin{equation}\label{length}
    \begin{cases}
        0 \le x_{l,m,n}\le \hat{L}, &\text{BS}\ l\ \text{satisfies}\ x_{l,n}=0,\\[-0.05cm]
        \displaystyle D_x-\hat{L} \le x_{l,m,n}\le D_x, &\text{BS}\ l\ \text{satisfies}\ x_{l,n}=D_x.\vspace{-0.2cm}
    \end{cases}
\end{equation}
where $\hat{L}$ denotes the length of waveguides.} Besides, the minimum gap between adjacent PAs are set as $\delta=\lambda_c$, where $\lambda_c$ is the wavelength of signal at frequency $f_c$. Thus, the PA deployment should also satisfy\vspace{-0.1cm}
\begin{equation}\label{gap}
        x_{l,m,n} - x_{l,m-1,n} \ge \delta,  \forall m \in \mathcal{M}/\{1\}.\vspace{-0.3cm}
\end{equation}
\subsection{Channel Model}
The channels in PASS assisted communication systems can be divided into two parts, namely the in-waveguide channels describe  the signal propagation inside the waveguide between its feed point to the PA, and the wireless channels describe the signal propagation between the PAs to users. The channel between the $n$-th waveguide in BS $l$ and user $k$ is \vspace{-0.1cm}
\begin{equation}\label{wireless}
\begin{aligned}
        \mathbf{g}_{l,n,k}=\left(2\kappa_c\right)^{-1}\left[\frac{e^{-j \kappa_c d_{l,1,n,k}}} {d_{l,1,n,k}},\cdots,\frac{e^{-j \kappa_c d_{l,M,n,k}}} {d_{l,M,n,k}}\right],
\end{aligned}\vspace{-0.1cm}
\end{equation}
where $\kappa_c$ is wave number at carrier frequency $f_c$. $d_{l,m,n,k}=\left\|\mathbf{L}_{l,m,n}-\mathbf{L}_{k}\right\|_2$ stands for the propagation distance between the $m$-th PA on the $n$-th waveguide in BS $l$ and user $k$. The wireless channel between BS $l$ and user $k$ can be given as\vspace{-0.1cm}
\begin{equation}\label{wireless_overall}
        \mathbf{g}_{l,k}= \left[\mathbf{g}_{l,1,k}^\mathrm{T}, \mathbf{g}_{l,2,k}^\mathrm{T}, \cdots, \mathbf{g}_{l,N,k}^\mathrm{T}\right]^\mathrm{T}\in\mathbb{C}^{MN\times 1}. \vspace{-0.1cm}  
\end{equation}
The channel between the $m$-th PA on the $n$-th waveguide and its feed point in BS $l$ can be given as\vspace{-0.1cm} 
\begin{equation}
        f_{l,m,n}=\rho_{l,m,n}e^{-j\kappa_gd_{l,m,n}},\vspace{-0.1cm}
\end{equation}
where $\rho_{l,m,n}$ is the power allocation factor of the corresponding PA. Assuming the equal power allocation strategy in this work, we have $\rho_{l,m,n}=1/\sqrt{M}$. $d_{l,m,n}=\left\|\mathbf{L}_{l,m,n}-\mathbf{L}_{l,n}\right\|_2$ stands for the in-waveguide propagation distance. $\kappa_g=n_e\kappa_c$ is the guided wave number with waveguide
effective refractive index $n_e$. Since each PA can be applied on one specific waveguide, this partially-connected architecture leads to the block-diagonal structure in in-waveguide channels. The in-waveguide channel of BS $l$ can be expressed as\vspace{-0.1cm}
\begin{equation}\label{in-waveguide}
\begin{aligned}
\mathbf{F}_l=\text{blkdiag}\left(\mathbf{f}_{l,1},\mathbf{f}_{l,2},\cdots,\mathbf{f}_{l,N}\right)^\mathrm{T}\in\mathbb{C}^{N\times MN},
\end{aligned}\vspace{-0.1cm}
\end{equation}
where $\mathbf{f}_{l,n}=\left[f_{l,1,n},f_{l,2,n},\cdots,f_{l,M,n}\right]^\mathrm{T}$. Combining the wireless channel~\eqref{wireless_overall} and in-waveguide channel~\eqref{in-waveguide}, the channel between BS $l$ and user $k$ can be expressed as \vspace{-0.1cm}
\begin{equation}
        \mathbf{h}_{l,k}=\mathbf{F}_l\mathbf{g}_{l,k} \in\mathbb{C}^{N\times 1}. \vspace{-0.4cm}       
\end{equation}

\subsection{Signal Model}
The signal received at user $k$ can be expressed as\vspace{-0.1cm}
\begin{equation}
\begin{aligned}
        y_k\!=\!\sum\nolimits_{l=1}^L\!\mathbf{h}_{l,k}^\mathrm{H}\mathbf{W}_{l}\mathbf{s}\!+\!n_k\!=\!\sum\nolimits_{l=1}^L\!\mathbf{h}_{l,k}^\mathrm{H}\!\sum\nolimits_{i=1}^K\!\mathbf{w}_{li}{s}_i\!+\!n_k,
\end{aligned}\vspace{-0.1cm}
\end{equation}
where $\mathbf{W}_{l}=\left[\mathbf{w}_{l1},\mathbf{w}_{l2},\cdots,\mathbf{w}_{lK}\right]\in\mathbb{C}^{N\times K}$ denotes the transmitting beamforming matrix of BS $l$, with $\mathbf{w}_{lk}$ standing for the beamformer allocated to user $k$ by BS $l$. $\mathbf{s}=\left[{s}_1,{s}_2,\cdots,{s}_K\right]^\mathrm{T}\in\mathbb{C}^{K\times 1}$ is the normalized information bearing signal vector with ${s}_k$ denoting the signal intended for use $k$ and satisfying $\mathbb{E}\left({s}_i{s}_j^*\right)=\delta_{i,j}$. $n_k$ is the additive white Gaussian noise at the receiver of user $k$ with power $\sigma^2$. Define $\mathbf{h}_{k}= \big[\mathbf{h}_{1,k}^\mathrm{T}, \mathbf{h}_{2,k}^\mathrm{T}, \cdots, \mathbf{h}_{L,k}^\mathrm{T}\big]^\mathrm{T}$ and $\mathbf{w}_{k}= \big[\mathbf{w}_{1,k}^\mathrm{T}, \mathbf{w}_{2,k}^\mathrm{T}, \cdots, \mathbf{w}_{L,k}^\mathrm{T}\big]^\mathrm{T}$. The signal received at user $k$ can be simplified as \vspace{-0.1cm}
\begin{equation}
        y_k=\mathbf{h}_{k}^\mathrm{H}\mathbf{W}\mathbf{s}+n_k=\mathbf{h}_{k}^\mathrm{H}\mathbf{w}_{k}{s}_k+\mathbf{h}_{k}^\mathrm{H}\sum\nolimits_{i\neq k}^K\mathbf{w}_{i}{s}_i+n_k,  \vspace{-0.1cm}      
\end{equation}
where $\mathbf{W}=\left[\mathbf{w}_{1},\mathbf{w}_{2},\cdots,\mathbf{w}_{K}\right]\in\mathbb{C}^{LN\times K}$ denotes the overall transmitting beamforming matrix of the cell-free system. The communication rate of user $k$ can be expressed as\vspace{-0.1cm}
\begin{equation}
      R_k=\log_2 \Big(1+\frac{\left|\mathbf{h}_{k}^\mathrm{H}\mathbf{w}_{k}\right|^2}{\sum_{i\neq k}^K\left|\mathbf{h}_{k}^\mathrm{H}\mathbf{w}_{i}\right|^2+\sigma^2}\Big).\vspace{-0.4cm}
\end{equation}


\subsection{Problem Formulation}
In this work, the digital beamforming design and pinching beamforming design of the proposed PASS assisted cell-free system are optimized aiming at maximizing the user sum rate under power limitation  and the PA deployment limitation of each BS. This optimization problem can be expressed as \vspace{-0.1cm}
\begin{subequations}\label{problem:sum_rate}
    \begin{align} 
        \label{obj}       
        \max_{\mathbf{W}, \left\{\mathbf{X}_{l}\right\}_{l=1}^L} \  &  \sum\nolimits_{k=1}^K R_k \\[-0.05cm]
        \label{constraint:pa}
        \mathrm{s.t.} \  & \left[\mathbf{X}_l\right]_{mn}\! \in\! \mathcal{X}_{l,m,n}, \forall l \!\in\! \mathcal{L}, \forall m \!\in\! \mathcal{M}, \forall n \!\in\! \mathcal{N}, \\[-0.05cm]
        \label{constraint:power}
        & \text{tr}(\mathbf{W}_{l}^\mathrm{H}\mathbf{W}_{l}) \le P_l, \forall l \in \mathcal{L},
    \end{align}\vspace{-0.5cm}
\end{subequations}

\noindent {where~\eqref{constraint:pa} is the PA deployment constraint, with $\mathbf{X}_l\in \mathbb{C}^{M\times N}$ denoting the PA deployment matrix for BS $l$. The element on row $m$ and column $n$ of $\mathbf{X}_l$ is $x_{l,m,n}$. $\mathcal{X}_{l,m,n}$ is the set of the values of $x_{l,m,n}$ which satisfy PA deployment constraints~\eqref{length} and~\eqref{gap}.} Constraint~\eqref{constraint:power} is the power budget constraint for all BSs, with $P_l$ denoting the available power of BS $l$. 
\vspace{-0.3cm}
\section{Sum Rate Maximization Design for PASS Assisted Cell-Free Systems}\vspace{-0.1cm}
\subsection{Problem Reformulation}
To tackle the non-convex objective function in problem~\eqref{problem:sum_rate}, the weighted minimum mean square error (WMMSE) method is adopted to transform~\eqref{obj} into a form that are convex in the following AO iterations~\cite{5756489}. Let $u_k \in \mathbb{C}$ denote the linear receive equalizer of user $k$, and $q_k > 0$ be the weight associated with the mean-square error (MSE). The MSE of user $k$ is \vspace{-0.1cm}
\begin{equation}
        \begin{aligned}
e_k 
&= \mathbb{E}\!\left[\,\big|u_k^* y_k - s_k\big|^2\,\right]=\\[-0.05cm]
& |u_k|^2 \!\Big(\sum\nolimits_{i=1}^K |\mathbf{h}_k^H \mathbf{w}_i|^2 + \sigma^2\Big)
   - 2\,\Re\!\left(u_k\,\mathbf{h}_k^H \mathbf{w}_k\right) + 1.
\end{aligned}\vspace{-0.1cm}
\end{equation}
The objective function~\eqref{obj} can be equivalent expressed as\vspace{-0.1cm}
\begin{equation}\label{wmmse}
        \min_{\{u_k,q_k\}} \ \sum\nolimits_{k=1}^K  \left( q_k e_k - \log q_k \right)-K.\vspace{-0.1cm}
\end{equation}
Then, problem~\eqref{problem:sum_rate} can be reformulated as \vspace{-0.1cm}
\begin{subequations}\label{problem:sum_rate_WMMSE}
    \begin{align}      
        \max_{\mathbf{W}, \left\{\mathbf{X}_{l}\right\}_{l=1}^L, \left\{u_{k}, q_{k}>0\right\}_{k=1}^K} \quad  &  \sum\nolimits_{k=1}^K \log q_k - q_k e_k \\[-0.05cm]
        \mathrm{s.t.} \quad  & \eqref{constraint:pa}, \eqref{constraint:power},
    \end{align}\vspace{-0.6cm}
\end{subequations}

\noindent The optimal equalizer $u_k$ and weight $q_k$ admit closed-form solutions:\vspace{-0.1cm}
\begin{equation}\label{uq}
        u_k^{\star} = \frac{\mathbf{h}_k^H \mathbf{w}_k}{\sum_{i=1}^K |\mathbf{h}_k^H \mathbf{w}_i|^2 + \sigma^2}, 
\quad 
q_k^{\star} = e_k^{-1}.\vspace{-0.3cm}
\end{equation}



\subsection{Digital Beamforming}
With fixed equalizer $u_k$, weight $q_k, \forall k\in\mathcal{K}$, and PA deployment $\mathbf{X}_l, \forall l\in\mathcal{L}$, the sum rate problem~\eqref{problem:sum_rate_WMMSE} can be transformed to\vspace{-0.1cm}
\begin{subequations}\label{problem:sum_rate_w}
    \begin{align}        
        \min_{\mathbf{W}} \quad  &  \sum\nolimits_{k=1}^K  \mathbf{w}_k^\mathrm{H}\mathbf{A}\mathbf{w}_k-2\sum\nolimits_{k=1}^K\Re\left(\mathbf{a}_{k}^\mathrm{H}\mathbf{w}_{k}\right) \\[-0.05cm]
        \mathrm{s.t.} \quad  & \eqref{constraint:power},
    \end{align}\vspace{-0.6cm}
\end{subequations} 

\noindent where $\mathbf{a}_{k}=u_kq_k\mathbf{h}_{k}$, $\mathbf{A}=\sum_{i=1}^Kq_i\left|u_i\right|^2\mathbf{h}_{i}\mathbf{h}_{i}^\mathrm{H}$. Problem~\eqref{problem:sum_rate_w} is convex and can be effectively solved with CVX~\cite{grant2014cvx}. {The worst case complexity of solving this problem using the interior point method is $\mathcal{O}\left(L^{4.5}N^{4.5}K^{4.5}\right)$.} To cut down the complexity of digital beamforming, we propose an alternative BS digital beamforming algorithm where the beamforming matrix of each BS is sequentially optimized. 
Given beamforming matrix $\mathbf{W}_i, \forall i\neq l$, digital beamforming problem with respect to BS $l$ can be expressed as \vspace{-0.1cm}
\begin{subequations}\label{problem:sum_rate_W_l}
    \begin{align}        
        \min_{\mathbf{W}_l} \quad  &  \sum\nolimits_{k=1}^K  \mathbf{w}_{l,k}^\mathrm{H}\mathbf{A}_{l,l}\mathbf{w}_{l,k}+2\sum\nolimits_{k=1}^K\Re\left(\mathbf{b}_{l,k}^\mathrm{H}\mathbf{w}_{l,k}\right) \\
        \mathrm{s.t.} \quad  & \text{tr}(\mathbf{W}_{l}^\mathrm{H}\mathbf{W}_{l}) \le P_l,
    \end{align}\vspace{-0.6cm}
\end{subequations} 

\noindent where $\mathbf{A}_{i,j}=\sum_{t=1}^Kq_t\left|u_t\right|^2\mathbf{h}_{i,t}\mathbf{h}_{j,t}^\mathrm{H}$, $\mathbf{a}_{l,k}=\sum_{i\neq l}^L\mathbf{A}_{l,i}\mathbf{w}_{i,k}-u_kq_k\mathbf{h}_{l,k}$. Using the Lagrange multipliers method, the Lagrange function of problem~\eqref{problem:sum_rate_W_l} can be given as \vspace{-0.1cm}
\begin{equation}
\begin{aligned}
        \!L\!\left(\mathbf{W}_l,\lambda_{l}\right)\!=\!&\sum\nolimits_{k=1}^K\!\!\mathbf{w}_{l,k}^\mathrm{H}\mathbf{A}_{l,l}\mathbf{w}_{l,k}\!+\!2\sum\nolimits_{k=1}^K\!\!\Re\!\left(\mathbf{a}_{l,k}^\mathrm{H}\mathbf{w}_{l,k}\right)\\[-0.05cm]
        &+\lambda_{l}\Big(\sum\nolimits_{k=1}^K\mathbf{w}_{l,k}^\mathrm{H}\mathbf{w}_{l,k}-P_l\Big),
\end{aligned}\vspace{-0.1cm}
\end{equation}
where $\lambda_{l}\ge0$ is the Lagrange multiplier. The stationary condition in KKT conditions lead to \vspace{-0.1cm}
\begin{equation}
        \mathbf{w}_{l,k}\left(\lambda_{l}\right)=\left(\mathbf{A}_{l,l}+\lambda_{l}\mathbf{I}_N\right)^{-1}\mathbf{a}_{l,k},\vspace{-0.1cm}
\end{equation}
Based on the complementary condition in KKT conditions, the optimal beamformer for user $k$ in BS $l$ is\vspace{-0.1cm}
 \begin{equation}\label{mu=0}
        \mathbf{w}_{l,k}^{\star}=\mathbf{w}_{l,k}\left(0\right)=\mathbf{A}_{l,l}^{\dagger}\mathbf{a}_{l,k},\vspace{-0.1cm}
\end{equation}
if $\sum_{k=1}^K\mathbf{w}_{lk}\left(0\right)^\mathrm{H}\mathbf{w}_{lk}\left(0\right)\le P_l$. Otherwise, the optimal beamformer for user $k$ in BS $l$ is\vspace{-0.1cm}
\begin{equation}\label{muneq0}
        \mathbf{w}_{l,k}^{\star}=\left(\mathbf{A}_{l,l}+\lambda_{l}^{\star}\mathbf{I}_N\right)^{-1}\mathbf{a}_{l,k},\vspace{-0.1cm}
\end{equation}
where $\lambda_{l}^{\star}$ is the solution to $\sum_{k=1}^K\mathbf{w}_{lk}\left(0\right)^\mathrm{H}\mathbf{w}_{lk}\left(0\right)= P_l$, which can be solved with  one
dimensional search~\cite{5756489}. The complexity of solving problem~\eqref{problem:sum_rate_W_l} with the Lagrange multipliers method mainly stems from the matrix inversion for $\mathbf{A}_{l,l}+\lambda_{l}\mathbf{I}_N$, which is $\mathcal{O}\left(N^3\right)$. The proposed AO algorithm for solving problem~\eqref{problem:sum_rate_w} is summarized in \textbf{Algorithm~\ref{algorithm2}}. {The complexity of \textbf{Algorithm~\ref{algorithm2}} can be given as $O_{\ref{algorithm2}}=\mathcal{O}\left(\log\left(\frac{1}{\epsilon_{\ref{algorithm2}}}\right)LKN^3\right)$.}


\begin{algorithm}[t]
\caption{AO algorithm for solving problem~\eqref{problem:sum_rate_w}.}\label{algorithm2}
\begin{algorithmic}[1]
\STATE Initialize feasible digital beamforming matrix  ${\mathbf{W}}_l, \forall l$. Set the AO convergence tolerance $\epsilon_{\ref{algorithm2}} > 0$. 
\REPEAT
\FOR{$l=1$ to L}
\STATE Given ${\mathbf{W}}_i, \forall i\neq l$, update $\mathbf{W}_l$ by solving problem~\eqref{problem:sum_rate_W_l} with~\eqref{mu=0} or~\eqref{muneq0}.
\ENDFOR
\UNTIL the objective function value of problem~\eqref{problem:sum_rate_w} experiences a fractional increase smaller than~$\epsilon_{\ref{algorithm2}}$ 
\end{algorithmic}
\end{algorithm}
\vspace{-0.3cm}
\subsection{Pinching Beamforming}
With fixed equalizer $u_k$, weight $q_k, \forall k\in\mathcal{K}$, and digital beamforming matrix $\mathbf{W}$, the sum rate problem~\eqref{problem:sum_rate_WMMSE} can be transformed to \vspace{-0.1cm}
\begin{subequations}\label{problem:sum_rate_X}
    \begin{align}        
        \min_{\left\{\mathbf{X}_{l}\right\}_{l=1}^L} \quad  &  \sum\nolimits_{k=1}^K  \mathbf{h}_k^\mathrm{H}\mathbf{B}\mathbf{h}_k-2\sum\nolimits_{k=1}^K\Re\left(\mathbf{b}_{k}^\mathrm{H}\mathbf{h}_{k}\right) \\[-0.05cm]
        \mathrm{s.t.} \quad  & \eqref{constraint:pa},
    \end{align}\vspace{-0.6cm}
\end{subequations} 

\noindent where $\mathbf{b}_{k}=u_kq_k\mathbf{w}_{k}$, $\mathbf{B}=\sum_{i=1}^Kq_i\left|u_i\right|^2\mathbf{w}_{i}\mathbf{w}_{i}^\mathrm{H}$. Note that the channel between BS $l$ and user $k$ can be reformulated as \vspace{-0.1cm}
\begin{equation}
\begin{aligned}
    &\!\!\!\mathbf{h}_{l,k}\!=\!\mathbf{F}_{l}\mathbf{g}_{l,k}\!=\!\left[\mathbf{f}_{l,1}^\mathrm{T}\mathbf{h}_{l,1,k},\mathbf{f}_{l,2}^\mathrm{T}\mathbf{h}_{l,2,k},\cdots,\mathbf{f}_{l,N}^\mathrm{T}\mathbf{h}_{l,N,k}\right]^\mathrm{T}\!\!\!\!=\!\!\\[-0.05cm]
    &\!\!\!\sum_{m=1}^M\!\!\left[\left[\mathbf{f}_{l\!,1}\right]_m\!\left[\mathbf{h}_{l\!,1\!,k}\right]_m\!,\cdots\!,\left[\mathbf{f}_{l\!,N}\right]_m\!\left[\mathbf{h}_{l\!,N\!,k}\right]_m\right]^\mathrm{T}\!\!=\!\!\sum_{m=1}^M\!\mathbf{h}_{l,k,m},
\end{aligned}  \vspace{-0.1cm} 
\end{equation}
where $\mathbf{h}_{l,k,m}$ stands for the channel between the $m$-th PA on all $N$ waveguides of BS $l$ and user $k$. Then the channel between all BS and user $k$ can be reformulated as \vspace{-0.1cm} 
\begin{equation}
        \mathbf{h}_{k}=\sum\nolimits_{m=1}^M\left[\mathbf{h}_{1,k,m}^\mathrm{T}, \cdots, \mathbf{h}_{L,k,m}^\mathrm{T}\right]^\mathrm{T}=\sum\nolimits_{m=1}^M\mathbf{h}_{k,m},\vspace{-0.1cm} 
\end{equation}
where $\mathbf{h}_{k,m}$ stands for the channel between the $m$-th PA on all $N$ waveguides of all BS and user $k$. Define auxiliary matrix\vspace{-0.1cm} 
\begin{equation}
        \mathbf{Q}_m=\left[\mathbf{q}_{1,m},\mathbf{q}_{2,m},\cdots,\mathbf{q}_{K,m}\right],\vspace{-0.1cm}
\end{equation}
where $\mathbf{q}_{k,m}=\big[\mathbf{q}_{1,k,m}^\mathrm{T}, \mathbf{q}_{2,k,m}^\mathrm{T}, \cdots, \mathbf{q}_{L,k,m}^\mathrm{T}\big]^\mathrm{T}$.
Problem~\eqref{problem:sum_rate_X} can be reformulated as \vspace{-0.1cm}
\begin{subequations}\label{problem:sum_rate_X_Q}
    \begin{align}        
        \min_{\left\{\mathbf{X}_{l}\right\}_{l=1}^L,\left\{\mathbf{Q}_{m}\right\}_{m=1}^M} \quad  &  f\left(\mathbf{X}_{l},\mathbf{Q}_{m}\forall l,m\right) \\[-0.05cm]
        \mathrm{s.t.} \quad  & \mathbf{Q}_m=\mathbf{H}_m, \forall m,\\[-0.05cm] 
        &\eqref{constraint:pa},
    \end{align}\vspace{-0.5cm}
\end{subequations} 

\noindent where $f\left( \mathbf{X}_{l}, \mathbf{Q}_{m}, \forall l,m\right)=\sum_{k} (\sum_{m}\mathbf{q}_{k,m}^\mathrm{H})\mathbf{B}(\sum_{m}\mathbf{q}_{k,m})-2\sum_{k}\Re\left(\mathbf{b}_{k}^\mathrm{H}(\sum_{m}\mathbf{q}_{k,m})\right)$, $\mathbf{H}_m=\left[\mathbf{h}_{1,m},\cdots,\mathbf{h}_{K,m}\right]$. Using the penalty method, problem~\eqref{problem:sum_rate_X_Q} can be reformulated as \vspace{-0.1cm}
\begin{subequations}\label{problem:sum_rate_X_Q_penalty}
    \begin{align}        
        \min_{\left\{\mathbf{X}_{l}\right\}_{l=1}^L,\left\{\mathbf{Q}_{m}\right\}_{m=1}^M} \quad  &  g\left(\mathbf{X}_{l},\mathbf{Q}_{m},\forall l,m\right) \\[-0.05cm]
        \mathrm{s.t.} \quad  & \eqref{constraint:pa},
    \end{align}\vspace{-0.6cm}
\end{subequations} 

\noindent where $g\left( \mathbf{X}_{l}, \mathbf{Q}_{m}, \forall l,m\right)=\sum_{k=1}^K  (\sum_{m}\mathbf{q}_{k,m}^\mathrm{H})\mathbf{B}(\sum_{m}\mathbf{q}_{k,m})-2\sum_{k=1}^K\Re\left(\mathbf{b}_{k}^\mathrm{H}(\sum_{m}\mathbf{q}_{k,m})\right)+\frac{1}{\rho}\sum_{m=1}^M\|\mathbf{H}_m-\mathbf{Q}_m\|_F^2$. In the following, the AO method is adopted to sequentially optimize $\{\mathbf{Q}_{m}\}_{m=1}^M$ and $\{\mathbf{X}_{l}\}_{l=1}^L$. With fixed $\{\mathbf{X}_{l}\}_{l=1}^L$, problem~\eqref{problem:sum_rate_X_Q_penalty} can be reformulated as the following unconstrained problem\vspace{-0.1cm}
\begin{equation}\label{problem:sum_rate_X_Q_penalty_Qm}
\min_{\left\{\mathbf{Q}_{m}\right\}_{m=1}^M} \quad   g\left(\mathbf{X}_{l},\mathbf{Q}_{m},\forall l,m\right),\vspace{-0.1cm}
\end{equation}  
The first-order optimality condition of $g\left(\mathbf{X}_{l},\mathbf{Q}_{m},\forall l,m\right)$ with respect to $\mathbf{q}_{k,m}$ yields\vspace{-0.1cm}
\begin{equation}\label{qkm}
        \frac{\partial g}{\partial \mathbf{q}_{k,m}^*}=\mathbf{B}\sum\nolimits_{i=1}^M\mathbf{q}_{k,i}^{\star}-\mathbf{b}_k+\frac{1}{\rho}\big(\mathbf{q}_{k,m}^{\star}-\mathbf{h}_{k,m}\big)=\mathbf{0}.\vspace{-0.1cm}
\end{equation}
\begin{algorithm}[t]
\caption{Penalty algorithm for solving problem~\eqref{problem:sum_rate_X_Q}.}\label{algorithm3}
\begin{algorithmic}[1]
\STATE {Initialize feasible PA deployment $\mathbf{X}_l$, the convergence tolerance $\epsilon_{\ref{algorithm3}}^1,\epsilon_{\ref{algorithm3}}^2 > 0$, the penalty factor $\rho$,  and the penalty update factor $\tau<1$.}
\REPEAT
\REPEAT
\STATE Given ${\mathbf{X}}_l, \forall l$, update ${\mathbf{Q}}_m, \forall m$, with~\eqref{solution_qkm}.
\STATE   Given ${\mathbf{Q}}_m, \forall m$, update ${\mathbf{X}}_l, \forall l$, with the element-wise optimization. 
\UNTIL the objective function value of problem~\eqref{problem:sum_rate_X_Q_penalty} experiences a fractional increase smaller than~$\epsilon_{\ref{algorithm3}}^2$.
\STATE Reduce the penalty factor with $\rho = \tau\rho$. 
\UNTIL the constraint violation is below threshold $\epsilon_{\ref{algorithm3}}^1$.
\end{algorithmic}
\end{algorithm}
Base on the fact that $\sum_{m=1}^M{\partial g}/{\partial \mathbf{q}_{k,m}^*}=\mathbf{0}$, we can obtain\vspace{-0.1cm}
\begin{equation}\label{sum_qkm}
       \sum\nolimits_{m=1}^M\!\mathbf{q}_{k,m}^{\star} \!\!=\!\!\big(\rho M\mathbf{B}+\mathbf{I}_{LN}\big)^{-1}\!\big(\rho M\mathbf{b}_k\!+\!\sum\nolimits_{m=1}^M\!\mathbf{h}_{k,m}\big).\vspace{-0.1cm}
\end{equation}
Substituting~\eqref{sum_qkm} in~\eqref{qkm}, the optimal solution of $\mathbf{q}_{k,m}$ can be solved as\vspace{-0.2cm}
\begin{equation}\label{solution_qkm}
        \mathbf{q}_{k,m}^{\star}\!=\!\rho  \mathbf{b}_k+\mathbf{h}_{k,m}-\rho\mathbf{B}\big(\rho M\mathbf{B}+\mathbf{I}_{LN}\big)^{-1}\!\big(\rho M\mathbf{b}_k+\sum_{m=1}^M\mathbf{h}_{k,m}\big).\vspace{-0.1cm}
\end{equation}
The complexity of solving problem~\eqref{problem:sum_rate_X_Q_penalty_Qm} with the closed-form solution mainly stems from the matrix inversion for $\rho M\mathbf{B}+\mathbf{I}_{LN}$, which is $\mathcal{O}\left(N^3L^3\right)$. With fixed $\{\mathbf{Q}_{m}\}_{m=1}^M$, problem~\eqref{problem:sum_rate_X_Q_penalty} can be reformulated as\vspace{-0.1cm}
\begin{subequations}\label{problem:sum_rate_X_Q_penalty_Xl}
    \begin{align}        
        \min_{\left\{\mathbf{X}_{l}\right\}_{l=1}^L} \quad  &  \sum\nolimits_{m=1}^M\|\mathbf{H}_m-\mathbf{Q}_m\|_F^2 \\[-0.05cm]
        \mathrm{s.t.} \quad  & \eqref{constraint:pa}.
    \end{align} \vspace{-0.5cm}
\end{subequations} 

\noindent The element-wise optimization method is adopted to alternatively adjust the location of $x_{l,m,n}$ while keeping other $LMN-1$ PAs fixed. The optimization problem with respect to $x_{l,m,n}$ is \vspace{-0.2cm} 
\begin{subequations}\label{problem:sum_rate_X_Q_penalty_xlmn}
    \begin{align}        
        \min_{x_{l,m,n}} \quad  &  \sum\nolimits_{k=1}^K\|\mathbf{h}_{l,k,m}-\mathbf{q}_{l,k,m}\|_2^2 \\[-0.05cm]
        \mathrm{s.t.} \quad  & \eqref{constraint:pa}.
    \end{align}\vspace{-0.5cm}
\end{subequations} 

\noindent Problem~\eqref{problem:sum_rate_X_Q_penalty_xlmn} can be solved with one dimensional search. The alternative element-wise optimization has a complexity of $\mathcal{O}\left(IQLMKN^2\right)$, where $Q$ and $I$ are the number of one dimensional search grid and the number of element-wise iterations, respectively.  The proposed penalty algorithm for solving problem~\eqref{problem:sum_rate_X_Q_penalty} is summarized in \textbf{Algorithm~\ref{algorithm3}}. {The complexity of \textbf{Algorithm~\ref{algorithm3}} can be given as $O_{\ref{algorithm3}}=\mathcal{O}\left(\log\big(\frac{1}{\epsilon^1_{\ref{algorithm3}}}\big)\log\big(\frac{1}{\epsilon^2_{\ref{algorithm3}}}\big)(L^3N^3+IQLMKN^2)\right)$.}
\subsection{Overall Algorithm}
\begin{algorithm}[t]
\caption{Overall AO Algorithm for solving problem~\eqref{problem:sum_rate_WMMSE}.}\label{algorithm1}
\begin{algorithmic}[1]
\STATE Initialize feasible digital beamforming matrix  ${\mathbf{W}}_l, \forall l$, and PA deployment ${\mathbf{X}}_l, \forall l$. Set the AO convergence tolerance $\epsilon_{\ref{algorithm1}} > 0$. 
\REPEAT
\STATE Given ${\mathbf{W}}_l$ and ${\mathbf{X}}_l, \forall l$, update $u_k$ and $q_k$, $\forall k$, with~\eqref{uq}.
\STATE Given ${\mathbf{X}}_l$, $u_k$, $q_k$, $\forall l,k$, update $\mathbf{W}_l$ using \textbf{Algorithm~\ref{algorithm2}}.
\STATE Given ${\mathbf{W}}_l$, $u_k$, $q_k$, $\forall l,k$, update $\mathbf{X}_l$ using \textbf{Algorithm~\ref{algorithm3}}.
\UNTIL the objective function value of problem~\eqref{problem:sum_rate_WMMSE} experiences a fractional increase smaller than~$\epsilon_{\ref{algorithm1}}$ 
\end{algorithmic}
\end{algorithm}
The overall AO algorithm for solving problem~\eqref{problem:sum_rate_WMMSE} is summarized in \textbf{Algorithm~\ref{algorithm1}}. {Deriving global optimality guarantees or explicit optimality gaps for the considered non-convex optimization problem is highly challenging. The proposed algorithm aims to obtain an efficient locally optimal solution. The proposed AO algorithm generates a non-decreasing sequence of objective values. As the mean square error in problem~\eqref{problem:sum_rate_WMMSE} is bounded below, \textbf{Algorithm~\ref{algorithm1}} is guaranteed to converge to a stationary point of problem~\eqref{problem:sum_rate_WMMSE}.  The complexity of \textbf{Algorithm~\ref{algorithm1}} can be given as $\mathcal{O}\left(\log\big(\frac{1}{\epsilon_{\ref{algorithm1}}}\big)(O_{\ref{algorithm2}}+O_{\ref{algorithm3}})\right)$.}
\begin{figure} [htbp]
\centering\vspace{-0.3cm}
\includegraphics[width=0.37\textwidth]{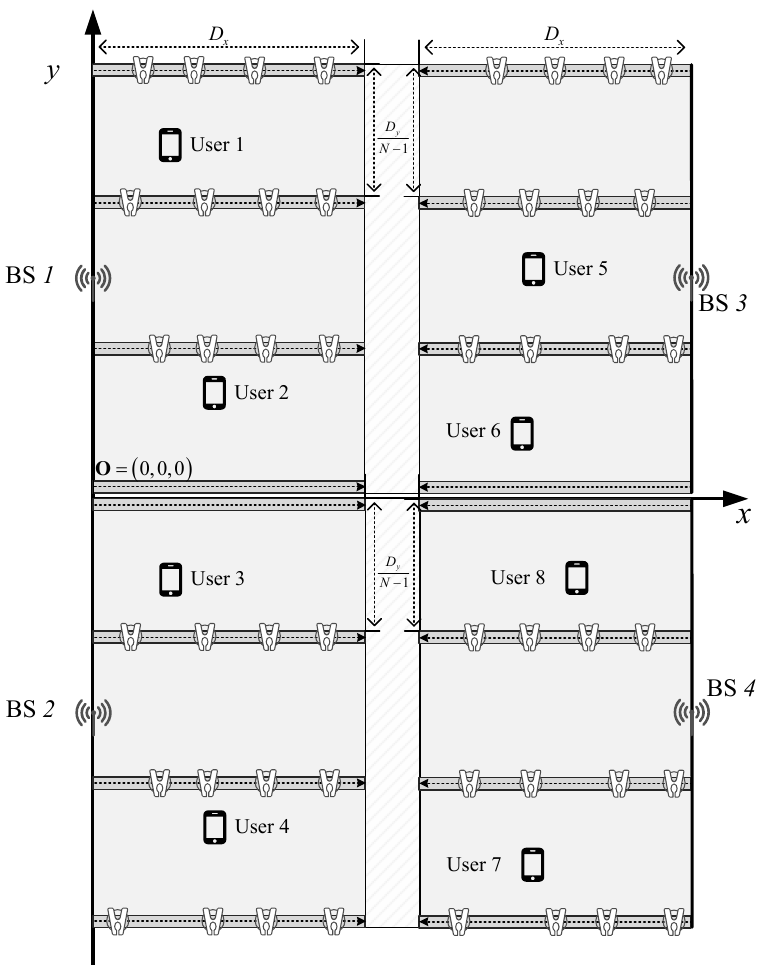}
 \caption{{Simulation setup for the PASS assisted cell-free communication system.}}\vspace{-0.4cm}
 \label{setup}
\end{figure}
\vspace{-0.3cm}
\section{Numerical Results}\label{Simulation}


This section introduces the simulation setups, benchmark schemes, and simulation results. {As depicted in Fig.~\ref{setup}, the number of BSs, the number of users, and  the number of waveguides in each BS are set as $L=4$, $K=8$, and $N=4$, respectively. The height of PASS is set as $d_h=3$ m. The service is spanned by $D_x\times D_y=30\ \text{m} \times 30\ \text{m}$. The length of waveguides are set as $\hat{L}=13$ m. The BS $1$, $2$, $3$, and $4$ are deployed at $(0,D_y/4,d_h)$ m, $(0,-D_y/4,d_h)$ m, $(D_x,D_y/4,d_h)$ m, and $(D_x,-D_y/4,d_h)$ m, respectively. The PA deployment constraint with respect to waveguide length for BS $1$ and $2$ is $0 \le x_{l,m,n}\le L$, while  that for BS $3$ and $4$ is $D_x-L \le x_{l,m,n}\le D_x$. All users are evenly divided between the coverage areas of four BSs, and their distribution is uniform within those areas.} The noise power at user receiver is set as $\sigma^2=-80$ dBm. The power budget at BS is set as $P_l=P=20$ dBm, $\forall l$. The waveguide effective refractive index is set as $n_e=1.4$. The carrier frequency is set as $f_c=28$ GHz. Unless otherwise specified, the parameter setup in following simulations are as decried above. Following benchmark schemes are compared to the proposed scheme

\begin{enumerate}
        \item \textbf{multiple-input and multiple-output (MIMO) Cell-Free:} The BS is equipped with $N$-antenna ULA. The resulted optimization problem can be solved by \textbf{Algorithm~\ref{algorithm2}} with $M=1$.
        \item \textbf{Uniformly Distributed PASS Assisted Cell-Free:} The PAs on each waveguide are uniformly distributed.
        \item \textbf{Discrete PASS Assisted Cell-Free:} The PAs on each waveguide can only be applied to $Z$ predefined sockets.
\end{enumerate}
In the simulation figures, legend ``C-PASS'' denotes the proposed PASS assisted cell-free system where the positions of the PAs can be continuously adjusted. Legend ``D-PASS'' denotes the proposed PASS assisted cell-free system where the PAs can only be attached to several predefined discrete positions. Legend ``U-PASS'' denotes the proposed PASS assisted cell-free system where the PAs are uniformly distributed along each waveguide.  Legend ``MIMO'' denotes the conventional MIMO enabled ISAC system.


\begin{figure} [htbp]
\centering\vspace{-0.3cm}
\includegraphics[width=0.5\textwidth]{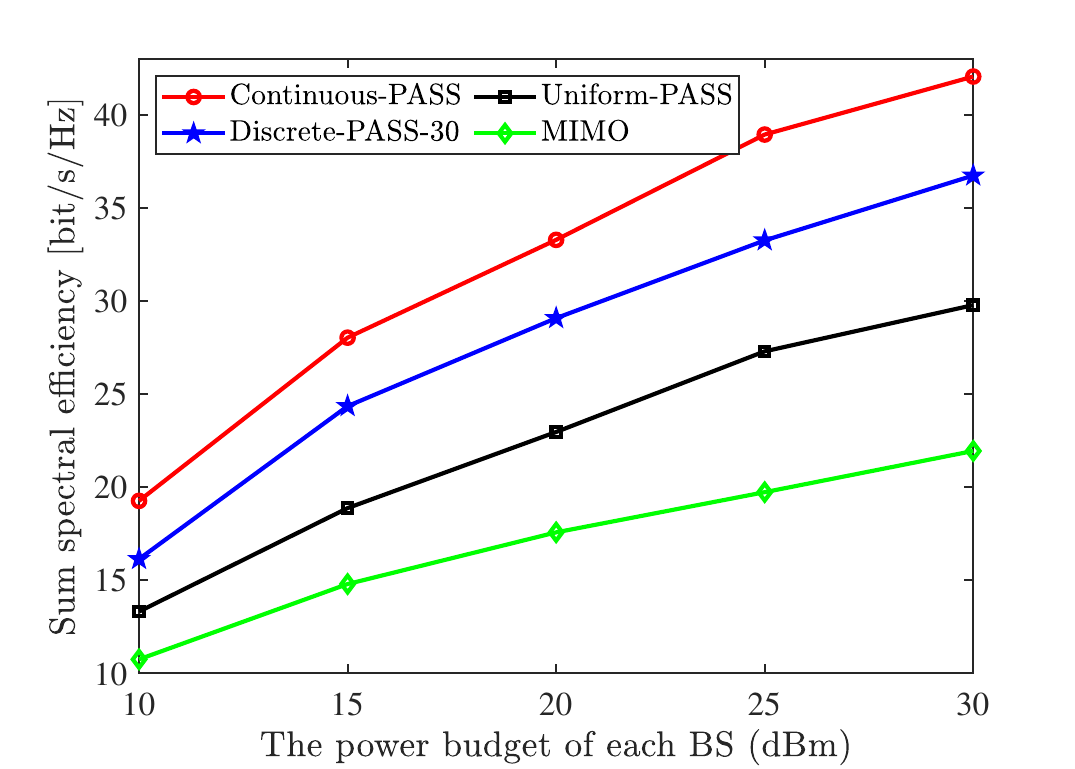}  
\caption{Sum spectrum efficiency versus the power budget of BS $P_l$.} 
\label{PASS_P} \vspace{-0.4cm}
\end{figure}
Fig.~\ref{PASS_P} illustrates the sum rate performance of the PASS enabled cell-free system and the benchmark schemes versus the power budget at each BS. To ensure a fair comparison between PASS based and conventional MIMO schemes, the waveguides in PASS are equipped with a single PA, i.e., $M=1$. As expected, the sum rate of all schemes increases with the BS transmit power, since higher transmit power improves the signal-to-interference-plus-noise ratio (SINR) at the users. Compared with benchmark schemes, the proposed PASS cell-free system consistently outperforms the MIMO cell-free system across all considered power budgets. {This performance gain stems from the unique features of the PASS architecture: pinching beamforming provides additional degrees of freedom for beamforming design, while the relocatable PAs can be positioned closer to users to reduce path loss. These advantages jointly contribute to the PASS gain observed in the PASS enabled cell-free systems.} Among the PASS cell-free schemes, the continuous PASS configuration achieves the best performance. By contrast, {the discrete PASS scheme suffers from a uniform performance loss due to its constrained PA deployment, which limits the degrees of freedom of pinching beamforming.} Similarly, the uniform PASS scheme is inferior to the optimized designs, {since its predefined PA deployment cannot adapt to the user distribution, leading to further performance degradation. }

\begin{figure} [htbp]
\centering\vspace{-0.3cm}
\includegraphics[width=0.5\textwidth]{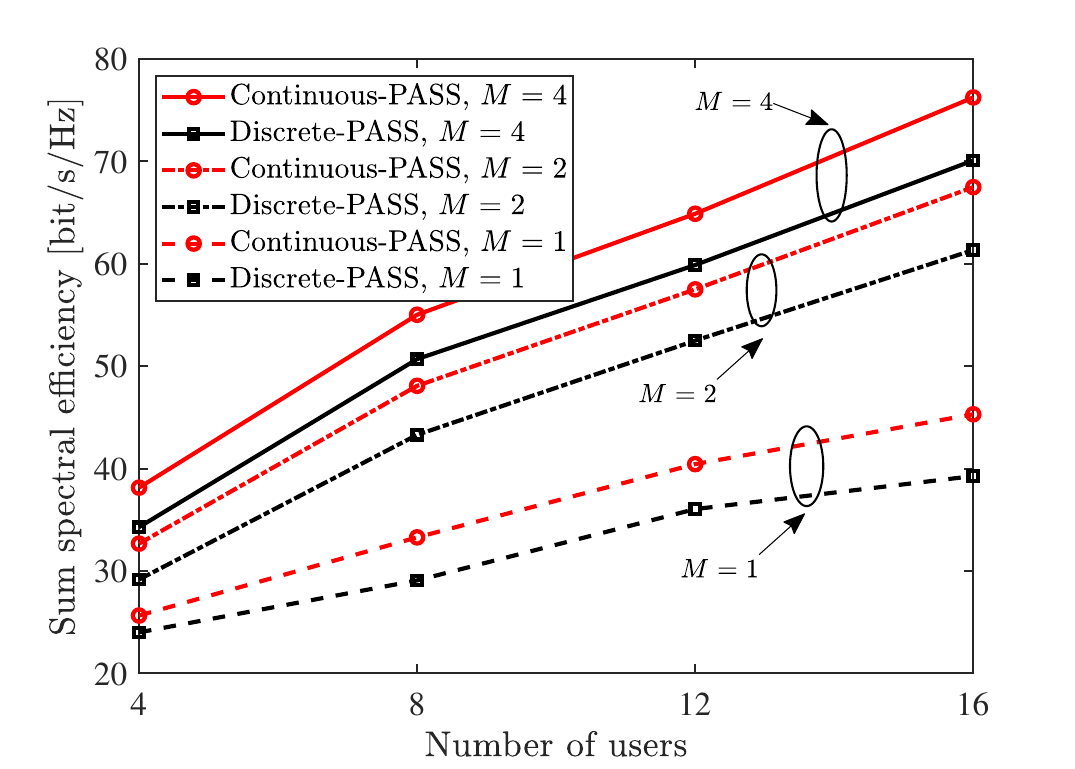} 
\caption{Sum spectrum efficiency versus the number of users $K$.} 
\label{PASS_K} \vspace{-0.4cm}
\end{figure}\begin{figure} [htbp]
\centering\vspace{-0.3cm}
\includegraphics[width=0.5\textwidth]{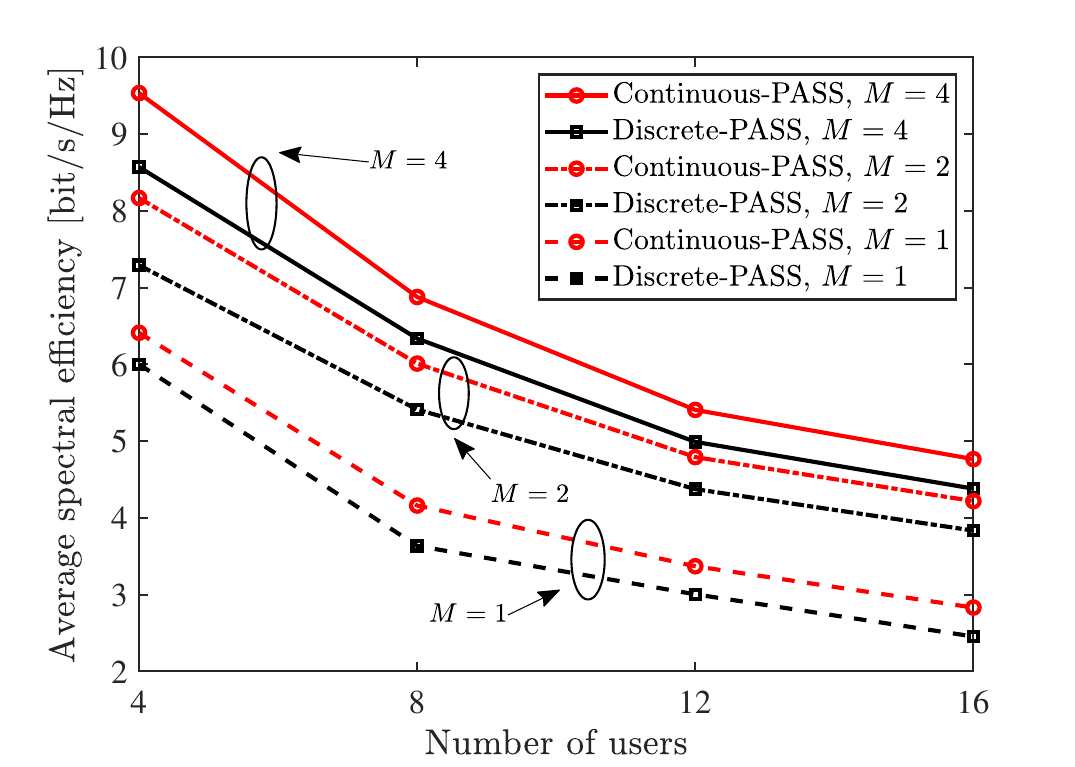} 
\caption{Average spectrum efficiency versus the number of users $K$.} 
\label{PASS_K_ave} \vspace{-0.4cm}
\end{figure} 

Fig.~\ref{PASS_K} and Fig.~\ref{PASS_K_ave} demonstrate the sum rate and average rate of the PASS cell-free system versus the number of users, respectively. {Owing to the user multiplexing capability provided by the joint digital and pinching beamforming design, the sum rate in Fig.~\ref{PASS_K} increases almost linearly with the number of users.} Moreover, as shown in Fig.~\ref{PASS_K_ave}, the cell-free architecture effectively alleviates the decline in average user rate as the number of users grows. This observation highlights two key insights. On the one hand, it demonstrates the advantage of joint multi-BS digital beamforming design in cell-free systems. On the other hand, it reveals that increasing the number of PAs per waveguide can further enhance system performance, {since more PAs provide additional degrees of freedom for pinching beamforming in PASS assisted cell-free systems\footnote{In practical systems, the design of $M$ should jointly consider communication performance, hardware complexity, and deployment cost, which remains an important future research direction. Moreover, due to the flexible and reconfigurable nature of PASS, a pragmatic design criterion for $M$ can also be determined in a trial-based or adaptive manner.}.}

\begin{figure} [htbp]
\centering\vspace{-0.3cm}
\includegraphics[width=0.5\textwidth]{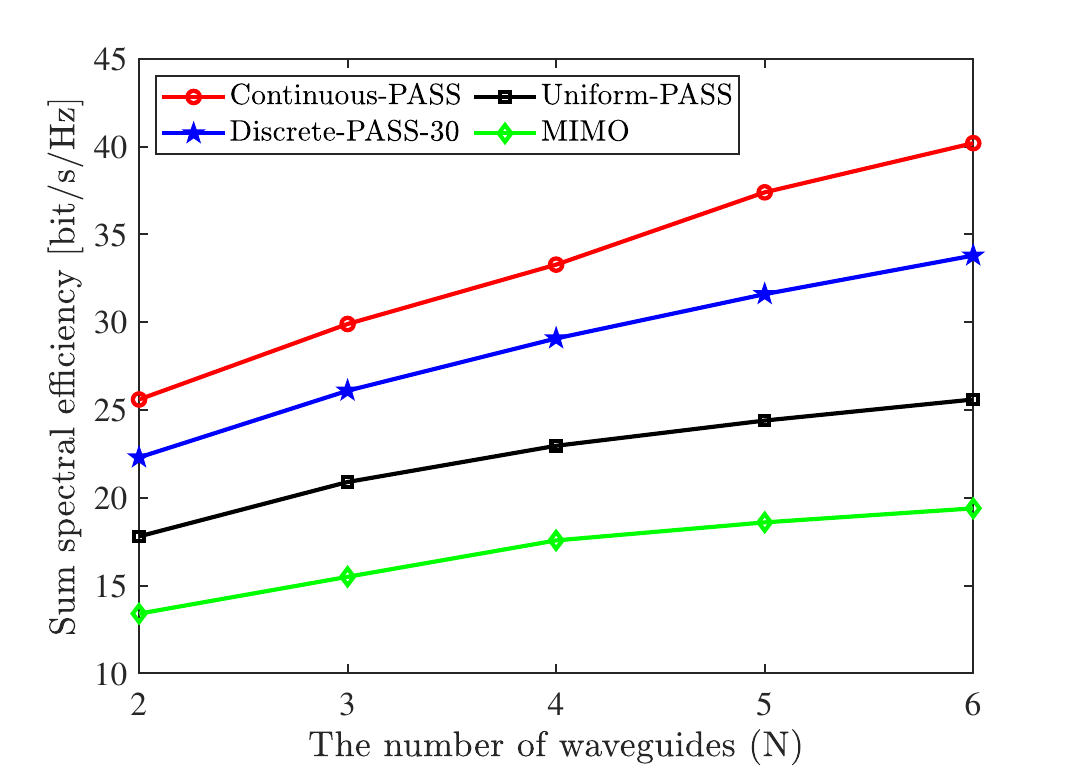} 
\caption{{Sum spectrum efficiency versus the number of waveguides $N$.}} 
\label{PASS_N}  \vspace{-0.4cm}
\end{figure}
{Fig.~\ref{PASS_N} illustrates the sum rate performance of the PASS enabled cell-free system and the benchmark schemes versus the number of waveguides at each BS. To ensure a fair comparison between PASS based and conventional MIMO schemes, the waveguides in PASS are equipped with a single PA, i.e., $M=1$. As the number of waveguides increases, the sum rate of all schemes improves due to the increased multiplexing gain brought by additional spatial transmission dimensions. Compared with conventional MIMO, PASS can further enhance the effective channel orthogonality through flexible antenna positioning and channel reconstruction, thereby utilizing the multiplexing gain more efficiently. Consequently, the proposed Continuous-PASS achieves the highest sum rate, followed by discrete PASS scheme and uniform PASS scheme.}

{Although the above simulations show that PASS assisted cell-free systems can outperform conventional cell-free MIMO systems, PASS may not always be preferable in practice. In relatively stable environments with limited blockage and slowly varying user distributions, carefully deployed fixed cell-free MIMO systems can already provide satisfactory coverage and spatial diversity with lower hardware complexity. In such cases, the additional flexibility of PASS may offer only marginal performance gains compared with its increased implementation complexity.}



\vspace{-0.3cm}
\section{Conclusions}\vspace{-0.1cm}
In this work,  a PASS assisted cell-free communication system is proposed. To address the formulated non-convex sum rate maximization  problem, an  AO algorithm was developed by leveraging the WMMSE method, the penalty method, and element-wise optimization. Simulation results demonstrated that the proposed PASS assisted cell-free system consistently outperforms benchmark schemes and further benefits from an increased number of PAs per waveguide. {In practical systems, waveguide attenuation, discrete PAs, and CSI feedback overhead may affect system performance. Incorporating these practical hardware constraints and developing low-overhead optimization schemes for cell-free PASS systems will be considered in our future work.}

\vspace{-0.2cm}
\bibliographystyle{IEEEtran}
\bibliography{myref}

\begin{thebibliography}{10}
\providecommand{\url}[1]{#1}
\csname url@samestyle\endcsname
\providecommand{\newblock}{\relax}
\providecommand{\bibinfo}[2]{#2}
\providecommand{\BIBentrySTDinterwordspacing}{\spaceskip=0pt\relax}
\providecommand{\BIBentryALTinterwordstretchfactor}{4}
\providecommand{\BIBentryALTinterwordspacing}{\spaceskip=\fontdimen2\font plus
\BIBentryALTinterwordstretchfactor\fontdimen3\font minus
  \fontdimen4\font\relax}
\providecommand{\BIBforeignlanguage}[2]{{%
\expandafter\ifx\csname l@#1\endcsname\relax
\typeout{** WARNING: IEEEtran.bst: No hyphenation pattern has been}%
\typeout{** loaded for the language `#1'. Using the pattern for}%
\typeout{** the default language instead.}%
\else
\language=\csname l@#1\endcsname
\fi
#2}}
\providecommand{\BIBdecl}{\relax}
\BIBdecl

\bibitem{11222687}
Z.~Yang, N.~Wang, Y.~Sun, Z.~Ding, R.~Schober, G.~K. Karagiannidis, V.~W. Wong,
  and O.~A. Dobre, ``Pinching antennas: Principles, applications and
  challenges,'' \emph{{IEEE} Wireless Commun.}, vol.~33, no.~2, pp. 175--184,
  2026.

\bibitem{11123791}
M.~Zeng, J.~Wang, G.~Zhou, F.~Fang, and X.~Wang, ``Energy-efficient design for
  downlink pinching-antenna systems with {QoS} guarantee,'' \emph{{IEEE} Trans.
  Veh. Technol.}, vol.~75, no.~2, pp. 3503--3507, 2026.

\bibitem{10945421}
Z.~Ding, R.~Schober, and H.~Vincent~Poor, ``Flexible-antenna systems: A
  pinching-antenna perspective,'' \emph{{IEEE} Trans. Commun.}, vol.~73,
  no.~10, pp. 9236--9253, 2025.

\bibitem{ngo2017cell}
H.~Q. Ngo, A.~Ashikhmin, H.~Yang, E.~G. Larsson, and T.~L. Marzetta,
  ``Cell-free massive {MIMO} versus small cells,'' \emph{{IEEE} Trans. Wireless
  Commun.}, vol.~16, no.~3, pp. 1834--1850, 2017.

\bibitem{bjornson2020scalable}
E.~Bj{\"o}rnson and L.~Sanguinetti, ``Scalable cell-free massive {MIMO}
  systems,'' \emph{{IEEE} Trans. Commun.}, vol.~68, no.~7, pp. 4247--4261,
  2020.

\bibitem{11018493}
H.~Wei, W.~Wang, W.~Ni, C.~Zhang, and Y.~Huang, ``Movable-antenna enabled
  cell-free networks,'' \emph{{IEEE} Trans. Veh. Technol.}, vol.~74, no.~10,
  pp. 16\,533--16\,537, 2025.

\bibitem{10967080}
T.~Han, Y.~Zhu, K.-K. Wong, G.~Zheng, and H.~Shin, ``Cell-free fluid antenna
  multiple access networks,'' \emph{{IEEE} Trans. Wireless Commun.}, vol.~24,
  no.~9, pp. 7237--7251, 2025.

\bibitem{11355743}
E.~Zhou, J.~Cui, Z.~Liu, Z.~Ding, and P.~Fan, ``Joint transmission for cellular
  networks with pinching antennas: System design and analysis,'' \emph{{IEEE}
  Trans. Wireless Commun.}, vol.~25, pp. 10\,175--10\,190, 2026.

\bibitem{5756489}
Q.~Shi, M.~Razaviyayn, Z.-Q. Luo, and C.~He, ``An iteratively weighted {MMSE}
  approach to distributed sum-utility maximization for a {MIMO} interfering
  broadcast channel,'' \emph{{IEEE} Trans. Signal Process.}, vol.~59, no.~9,
  pp. 4331--4340, 2011.

\bibitem{grant2014cvx}
M.~Grant and S.~Boyd, ``{CVX}: Matlab software for disciplined convex
  programming, version 2.1,'' [Online]. Available:\url{http://cvxr.com/cvx},
  2014.

\end{thebibliography}

\end{document}